\documentclass[preprint,pra,superscriptaddress,aps,showpacs,showkeys,amsmath,floatfix]{revtex4-1}

\usepackage{amsfonts}
\usepackage{amssymb}
\usepackage{graphics}
\usepackage{graphicx}
\usepackage{epstopdf}
\usepackage{dcolumn}
\usepackage{bm}
\usepackage{longtable}
\usepackage{epsfig}
\usepackage{times}
\usepackage{url}
\usepackage{color}

\begin{document}

\title{Dynamical control of nuclear isomer depletion via electron vortex beams}

%
%

\author{Yuanbin \surname{Wu}}
\affiliation{Max-Planck-Institut f\"ur Kernphysik, Saupfercheckweg 1, D-69117 Heidelberg, Germany}

\author{Simone Gargiulo}

\affiliation{Institute of Physics, Laboratory for Ultrafast Microscopy and Electron Scattering, \'Ecole Polytechnique F\'ed\'erale de Lausanne, Station 6, Lausanne 1015, Switzerland}

\author{Fabrizio Carbone}
\affiliation{Institute of Physics, Laboratory for Ultrafast Microscopy and Electron Scattering, \'Ecole Polytechnique F\'ed\'erale de Lausanne, Station 6, Lausanne 1015, Switzerland}

\author{Christoph H. \surname{Keitel}}
\affiliation{Max-Planck-Institut f\"ur Kernphysik, Saupfercheckweg 1, D-69117 Heidelberg, Germany}
\author{Adriana \surname{P\'alffy}}
\affiliation{Max-Planck-Institut f\"ur Kernphysik, Saupfercheckweg 1, D-69117 Heidelberg, Germany}
\affiliation{Department of Physics, Friedrich-Alexander-Universit\"at  Erlangen-N\"urnberg,  D-91058 Erlangen, Germany}

\date{\today}

\begin{abstract}
Long-lived excited states of atomic nuclei can act as energy traps. These states, known as nuclear isomers, can store a large amount of energy over long periods of time, 
with a very high energy-to-mass ratio.  Under natural conditions, the trapped energy is only slowly released, limited by the long isomer lifetimes. Dynamical external control of nuclear state population  has proven so far very challenging, despite ground-breaking incentives for a clean and efficient energy storage solution.
Here, we describe a protocol to achieve the external control of the isomeric nuclear decay  by using electrons whose wavefunction has been especially designed and reshaped  on demand. Recombination of these electrons into the atomic shell around the isomer can lead to the controlled release of the stored nuclear energy. On the example of  $^{93m}$Mo, we show that the use of tailored electron vortex beams increases the depletion by four orders of magnitude compared to the spontaneous nuclear  decay of the isomer. Furthermore, specific orbitals can sustain an enhancement of the recombination cross section for vortex electron beams by as much as six orders of magnitude, providing a handle for manipulating the capture mechanism. These findings open new prospects for controlling the interplay between atomic and nuclear degrees of freedom, with potential energy-related and high-energy radiation sources applications.

\end{abstract}

\maketitle


Nuclear isomers are metastable, long-lived excited states of atomic nuclei. Their direct decay to lower-lying levels is strongly suppressed, typically due to large differences in either spin, nuclear shape or  spin projection on the nuclear symmetry axis \cite{Walker1999, Walker2020}.
In some nuclei with an advantageous configuration of the nuclear excited states, an excitation to a level above the isomeric state (termed gateway state) can lead to the nuclear decay directly to a level below the isomer itself, thus reaching the ground state in a fast cascade.

\begin{widetext}
\begin{figure}[h]
\includegraphics[width=\textwidth]{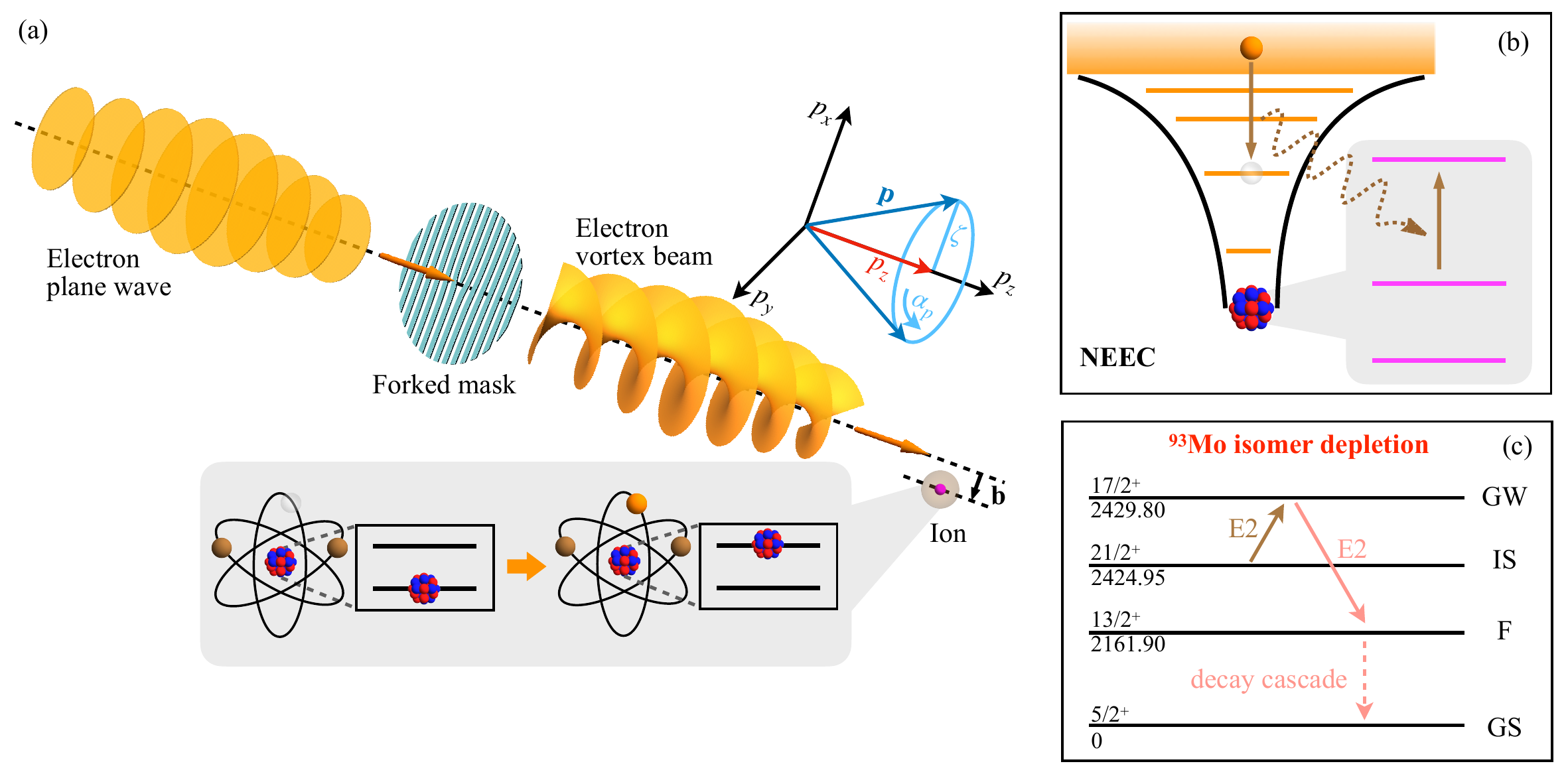}
\caption{\label{fig1} 
{\bf NEEC and isomer depletion with an electron vortex beam} 
(a) A plane-wave electron beam incident on a 
ed mask generates the electron vortex beam. Upon hitting on an ion beam with impact parameter $\mathbf{b}$, the electrons recombine into atomic vacancies. (b) At the resonant continuum electron energy, electron recombination (orange atomic shell levels on the left) will be accompanied by nuclear excitation (magenta nuclear states on the right) in the process of NEEC. (c) Partial level scheme of $^{93}$Mo.   The nuclear isomeric ($I$), gateway ($GW$), intermediate ($F$) and ground state ($GS$) levels are labeled by their spin, parity and energy in keV. The transitions $IS\rightarrow GW$ and $GW\rightarrow F$ are both of $E2$ type. Energy intervals are not to scale.
}
\end{figure}
\end{widetext}

Such a process is called isomer  depletion, since it allows for the depopulation of the isomeric state and thus a controlled release of the energy stored in the metastable nucleus. 
%
A typical example is the case of the 2425 keV $^{93m}$Mo isomer with a halflife of 6.8 h, for which we present the relevant partial level scheme in Fig.~\ref{fig1}. A 4.85 keV excitation from the isomer to the gateway state at 2430 keV should release the entire stored energy within only 4~ns. This appealing example has been often mentioned in the context of potential nuclear energy storage solutions  without involving fission or fusion \cite{Walker1999,GunstPRL2014,PalffyPRL2007,ChiaraNature2018}.

One of the most intriguing means to externally drive the transition from the isomer to the gateway state is via coupling to the atomic shell. In the process of nuclear excitation by electron capture (NEEC), an electron recombining into an atomic vacancy of an ion transfers resonantly its energy to the nucleus. The sum of the free electron energy and capture orbital binding energy must thereby match, within the uncertainty relations, the nuclear transition energy. This process, originally predicted in 1976 \cite{GoldanskiiPLB1976}, attracted a number of theoretical studies \cite{Cue1989,Kimball2,HarstonPRC1999,GosselinPRC2004,PalffyPRA2006} prior to the first claim of experimental observation in 2018 \cite{ChiaraNature2018}. Interestingly, the NEEC experiment was investigating exactly the isomer depletion transition in $^{93}$Mo. As theoretical works contradict the experimental results \cite{Wu2019,Polasik2021}, the subject is at present a matter of vivid debate \cite{Guo2021,Chiara2021}. Controversy aside, the overall consensus is that due to the small nuclear transition energy to the gateway state of $^{93m}$Mo, NEEC should be stronger than photoexcitation.

So far, the NEEC process has been considered for the case of plane-wave electrons  captured by ions which are initially in their electronic ground state. However, few recent works suggested that the NEEC cross section can be influenced by the ion's out of equilibrium conditions \cite{WuPRA2019,Gargiulo2020} or a different shape of the electronic wave function \cite{Madan2020}. Here, we take an important step to investigate the process of NEEC considering specially designed electron beams, which are tailored to enhance the nuclear excitation. Our results show that capturing an electron with a  properly reshaped  wavefunction can lead to an increase of the NEEC cross section by few orders of magnitude, depending on the specific situation considered. 

In recent years, the  achieved capability to fabricate phase masks with nanometer precision rendered possible to control the coherent superposition of matter waves producing typical interference patterns by spatial reshaping of a particle’s wave-function \cite{UchidaNature2010,VerbeeckNature2010,McMorranScience2010,ClarkNature2015,Luski2021}. Particularly interesting is the case of so-called vortex beams, which consist of a stream of particles whose wavefunction spatial profile has been modulated to become chiral and carry an orbital angular momentum.  

Optical vortices have been studied in the context of quantum communications, nano-plasmonics and optical trapping \cite{ShenLSA2019, BliokhPR2015}, while imparting chirality to massive composed particles has been proposed as a method to study \cite{LloydRMP2017,BliokhPR2017,VanacoreNM2019,zhao2021decay} and even manipulate \cite{LarocqueNP2018, ClarkNature2015, KaminerNP2015, Madan2020} the inner structure of neutrons, protons, ions and molecules. 
Electron vortex beams carry both orbital angular momentum about their beam axis and the electron's intrinsic spin momentum. Experimentally, they are produced by a number of techniques such as phase-plates, holographic gratings, magnetic monopole fields or chiral plasmonic near fields \cite{LloydRMP2017, BliokhPR2017, UchidaNature2010, VerbeeckNature2010, McMorranScience2010, VanacoreNM2019}, with angular momenta of up to $1000$ $\hbar$ already demonstrated. The angular momentum aspect is particularly important for nuclear transitions which display in the low-energy region mostly a dipole-forbidden character. The transition multipolarity, for instance, electric quadrupole ($E2$) or magnetic dipole ($M1$), together with the recombination orbital, impose strict selection rules on which angular momentum components of the incoming electron beam will undergo NEEC. While plane wave electron beams have a fixed  partial wave expansion in all multipoles, vortex beams can be shaped on purpose to enhance and control the NEEC outcome. 



A possible experimental implementation of this idea is depicted in Fig.~\ref{fig1}(a).  A plane wave electron beam is incident  on a phase mask which reshapes the wavefunction generating an electron vortex beam. We illustrate here a so-called forked mask as an example. The vortex beam 
 is incident on ions with atomic vacancies that facilitate the NEEC process. The electron energy is chosen such as to match resonantly the nuclear transition energy upon recombination into a chosen orbital as shown in  Fig.~\ref{fig1}(b). As examples we consider the canonical case of $^{93}$Mo, whose partial level scheme is depicted in 
Fig.~\ref{fig1}(c). The NEEC transition between the isomer and gateway states has 4.85 keV and $E2$ multipolarity. A second example envisaging a 19.70 keV $M1$ transition 
from the $^{152m}$Eu isomer at 45.60 keV isomer to a gateway state will also be considered. These examples are generic, and were chosen to demonstrate the effect on the two most frequently occurring nuclear transition  multipolarities ($E2$ and $M1$) in the energy range relevant for NEEC. For a plane-wave electron beam, the maximal NEEC cross section for depletion of $^{93m}$Mo occurs for recombination into the $2p_{3/2}$ orbital of a Mo$^{36+}$ ion \cite{WuPRL2018,GunstPRE2018}. This charge state is sufficient for providing the maximum number of vacancies in the $2p_{3/2}$ orbital. On the other hand, it ensures that the NEEC channel is allowed, with the resonance continuum electron energy of only approx. 52 eV. A higher charge state would close the NEEC channel due to the slight increase of electronic binding energies.

We consider a vortex beam with the longitudinal linear momentum $p_z$,  the modulus of the transverse momentum $|{\bf{p}}_{\bot}| = \zeta$, and the topological vortex charge, a quantity related to the electron orbital angular momentum, denoted by $m$  \cite{BliokhPR2017, BliokhPRL2011}. The corresponding electron wave function can be written as
\begin{equation}
  \psi_s ({\bf{r}}) = \int \frac{d^2 {\bf{p}}_{\bot}}{(2\pi)^2} a_{\zeta m}({\bf{p}}_{\bot}) u_{{\bf{p}}s} e^{i {\bf{p}} \cdot {\bf{r}}},
\end{equation}
where $a_{\zeta m}({\bf{p}}_{\bot}) = (-i)^m e^{i m \alpha_p} \delta(|{\bf{p}}_{\bot}| - \zeta) / \zeta$ and $u_{{\bf{p}}s}$ is the electron bispinor which corresponds to the plane-wave solution with momentum $\bf{p}$ and the spin state $s$. The linear momenta of the plane-wave components are given by ${\bf{p}} = ({\bf{p}}_{\bot}, p_z) = (\zeta \cos{\alpha_p}, \zeta \sin{\alpha_p}, p_z)$, as sketched in Fig.~\ref{fig1}. We choose the $Oz$ axis  parallel to the incident electron beam. To specify the lateral position of the ion with regard to the central axis of the incident electron beam, the impact parameter ${\bf{b}}$ is introduced \cite{BliokhPR2017, SerboPRA2015}. The advantage of the vortex beam comes into play when restricting the impact parameter \cite{BliokhPR2017, SerboPRA2015}. Otherwise, an average over arbitrary  impact parameters in the entire beam range will limit the enhancement factor for the NEEC rate to a factor $p/p_z$. We therefore restrict the impact parameter region to $|{\bf{b}}| \leqslant b$, with $b$ chosen accordingly as a function of the incoming electron momentum. The incident electron current is averaged over the impact parameter region.


In order to calculate the NEEC cross sections, the vortex beam is mapped upon the partial wave expansion of the continuum electron wave function. The resulting NEEC rate $Y_{neec}^{i\rightarrow g}$ can be written as a function of the reduced transition probability for the nuclear transition, electronic radial wave function integrals, and the vortex beam parameters $m$, $\zeta$ and $\alpha_p$ (see Methods). The total NEEC cross section can be written as a function of the continuum electron energy $E$, 
\begin{equation}
  \sigma_{neec}^{i\rightarrow g}(E) = \frac{4 \pi^2}{pJ_z}  Y_{neec}^{i \rightarrow g} \mathcal{L}(E-E_0),
\end{equation}
where $p$ is the modulus of the continuum electron momentum, $J_z$ is the total incident current which can be calculated via Ref. \cite{BliokhPRL2011}, and $\mathcal{L}(E-E_0)$ a Lorentz profile centered on the resonance energy $E_0$ and with a full width half maximum given by the width of the nuclear excited state. Typically, the nuclear widths are very narrow (for example, $\Gamma_g=10^{-7}$ eV for the case of 
$^{93m}$Mo), such that $\mathcal{L}(E-E_0)$ is approximated with  a Dirac-deltalike profile. Integrating over the continuum electron energy, we obtain the so-called resonance strength $S_v$. We compare this value with the resonance strength $S_p$ obtained for the case of a plane wave electron beam. 

\begin{figure}[!h]
\centering
\includegraphics[width=0.45\linewidth]{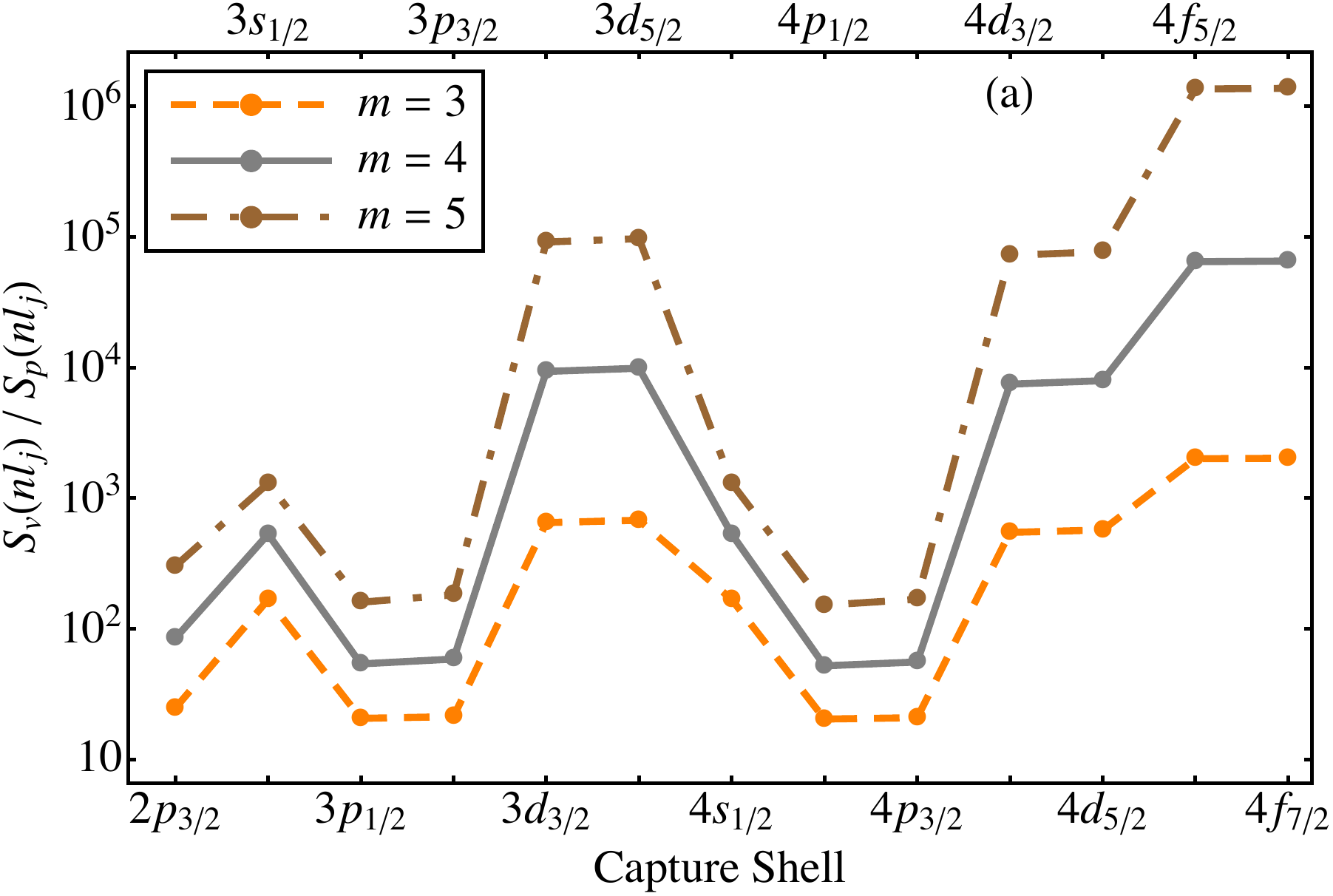}%
\hspace{0.3cm}
\includegraphics[width=0.5\linewidth]{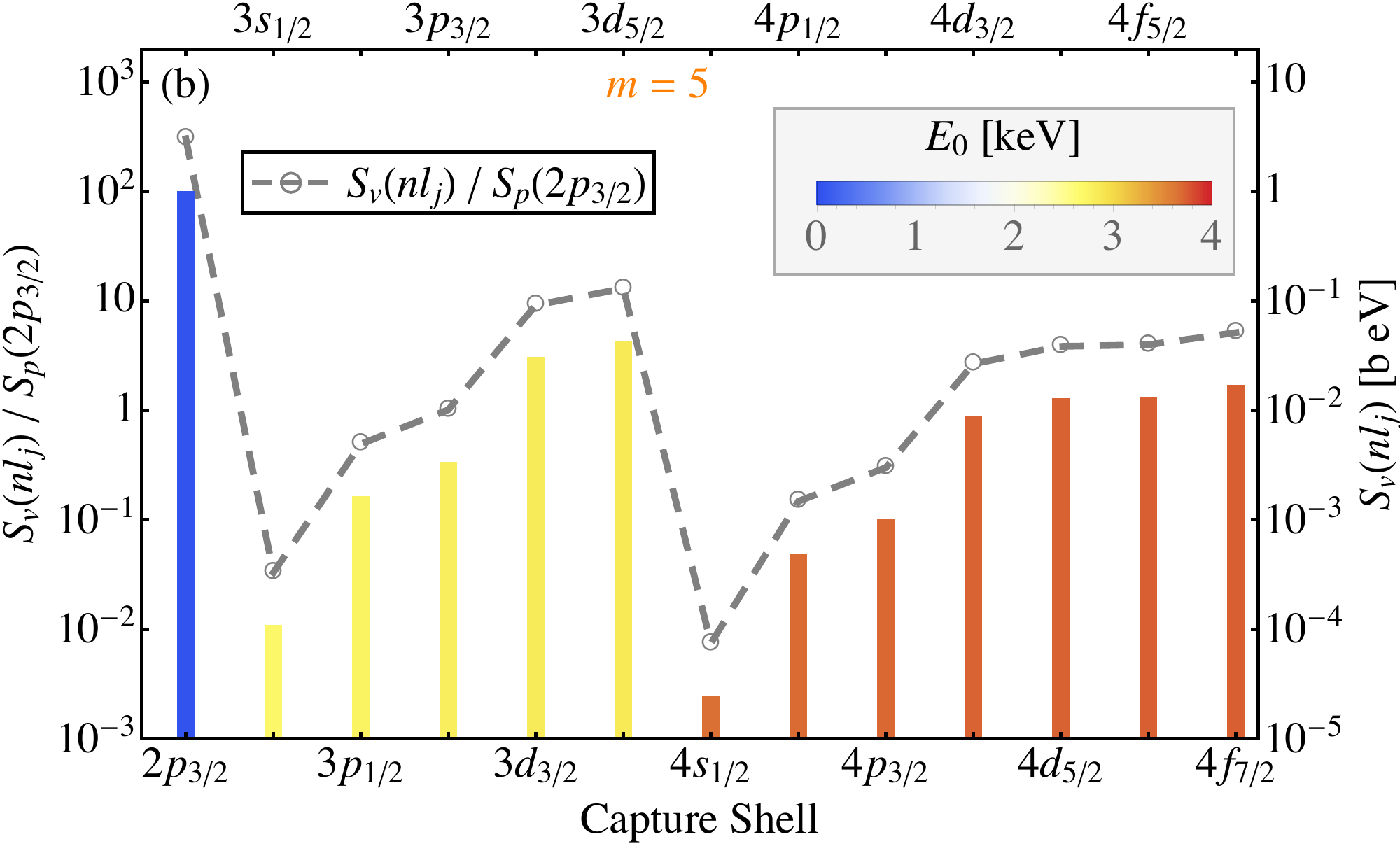}%

\caption{\textbf{NEEC integrated cross section enhancement for the $4.85$ keV  nuclear transition depleting $^{93m}$Mo.} (a) The enhancement ratio $S_v(nl_j)/S_p(nl_j)$ comparing vortex and plane wave electron beams for recombination orbitals in the range $2p_{3/2}$ to $4f_{7/2}$. (b) The ratio  $S_v(nl_j)/S_p(2p_{3/2})$ of vortex beam versus maximal plane wave NEEC resonance strengths corresponding to recombination into the $2p_{3/2}$ orbital (left-hand axis, grey dashed curve with circle), and the absolute values of $S_v(nl_j)$ (right-hand axis, vertical colored bars). We consider three values of the topological charge $m=3,\, 4,\, 5$ (a) or just $m=5$ (b), with $\zeta = p_z$ and impact parameter range $\zeta b = 1$. The resonant electron energy $E_0
$ is presented in color coding. 
\label{fig:mo93v}}
\end{figure}

 We focus our attention first to the case of  $^{93m}$Mo and electron recombination into the ground state of the Mo$^{36+}$ ion. We consider NEEC into the ground state configuration of the Mo$^{36+}$ ion into orbitals ranging from $2p_{3/2}$ to $4f_{7/2}$. The  continuum electron resonance energy for recombination into $2p_{3/2}$ is $52$ eV, while for the higher shell orbitals the values lie  between $2.7$ keV and $2.9$ keV for the $M$ shell and  between $3.6$ keV and $3.8$ keV for the $N$ shell.
The ratio $S_v/S_p$ as a function of the capture orbital for three values of  topological charge $m=3,\, 4, \, 5$ is presented in Fig.~\ref{fig:mo93v}(a). The vortex beam parameters are chosen such that $\zeta = p_z$ for the impact parameter range $b=1/\zeta$. Figure \ref{fig:mo93v}(a) shows that, depending on the recombination orbital, the tailored vortex electron beam leads to an enhancement between two ($p$~orbitals) and six orders of magnitude ($f$~orbitals) in the NEEC resonance strength. Although the enhancement for the capture into $M$- and $N$-shell orbitals is impressive,  these are not the capture orbitals with the largest cross section.

Provided that atomic vacancies are available, NEEC into the $2p_{3/2}$ is the most efficient isomer depletion channel.
For  an incident vortex beam, the resonance strength for NEEC into this orbital is increased by two orders of magnitude as compared to the  plane wave electron beams so far considered in the literature.
 This is demonstrated in Fig.~\ref{fig:mo93v}(b) which shows the vortex beam resonance strength scaled by the maximum value reached for a plane wave setup. In the vortex beam setup, also NEEC into the $3d$ or $4d$ and $4f$ orbitals exceeds the plane wave value for recombination into $2p_{3/2}$, however only by one order of magnitude. Still, this might become advantageous to ease the charge state requirements, or when the continuum electron energy cannot be decreased to very small energies.

 Angular momentum conservation in the NEEC process imposes selection rules for the continuum electron partial wave (see Methods) as a function of recombination orbital and nuclear transition multipolarity. These selection rules reflect also upon and determine the most efficient vortex charge $m$ for a particular NEEC process. 
 For instance a vortex beam with $m>5$ would further increase  NEEC into $d$ and $f$ orbitals. However, increasing $m$ at values above $m=5$ has less further enhancement effect on the NEEC resonance strength for the $2p_{3/2}$ orbitals. Depending on the envisaged electron beam energy (and therefore capture orbital), the proper choice of vortex beam topological charge $m$ can maximize the NEEC resonance strength. The new aspect here, specifically related to vortex beams,   is that  $m$ acts as a new degree of freedom and can be dynamically controlled   on an ultrafast timescale, as detailed below.

We now turn to a different example which investigates NEEC for a $M1$ nuclear transition in  $^{152}$Eu. This isotope has an isomer with 9.3 h halflife lying 45.60 keV above the ground state. The envisaged gateway state lies at 65.30 keV and is connected by an $M1$ transition to the isomer. Once the gateway state is reached, the nucleus will decay within approx. 1~$\mu$s with a branching ratio of 0.42 to the ground state. For this case, we consider NEEC occuring into a bare Eu ion. Table~\ref{table:eu152v} displays the plane wave and vortex electron beam NEEC resonance strengths for the cases of $m=3$ and $m=5$, assuming $\zeta = p_z$ and  $\zeta b = 1$. 

The enhancements compared to the equivalent plane wave case are less dramatic, with factors between 1.4 and approx. 600. The lowest factor of 1.4 occurs in the case of NEEC into the $2s_{1/2}$ orbital and stems mainly from the factor  $p/p_z$.
However, the startling feature in the case of $^{152}$Eu is the ability to change the most efficient capture orbital. For an $M1$ transition, the strongest NEEC resonance strength for a plane wave electron beam occurs for the recombination into the lowest available $s$ orbital. For the specific case of $^{152}$Eu, with its nuclear transition and electronic binding energies, this would be the $2s$ orbital. Surprisingly, the tailored vortex beam changes this rule of thumb, as  the strongest NEEC occurs for the $2p_{1/2}$ orbital (for $m=3$) or for the $2p_{3/2}$ orbital ($m=5$). 
Thus, by manipulating the wavefunction of the incident electronic beam, it is possible not only to enhance rates but also to shift the maximum effect between orbitals.

In view of the many methods developed to produce specific atomic vacancies \cite{rudek2012ultra,SteckPPNP2020}, this result can have important consequences for our ability to manipulate the nuclear excitation. Vortex beam angular momentum, electron energy and atomic vacancies can be dynamically and simultaneously controlled to optimize isomer depletion. In fact, the topological charge of the vortex beam  impinging on the isomers,  i.e., the value of $m$, can be switched dynamically on an ultrafast timescale by modulating the properties of plasmonic  \cite{VanacoreNM2019, kim2010synthesis, wang2019dynamical} and light phase masks  \cite{Lembessis2014,Lembessis2017}. Also when using 
physical phase plates such as the forked mask in Fig.~\ref{fig1}, deflector coils or apertures can select the desired vortex topological charge \cite{PohlSR2017}. 
With such dynamical control to optimize isomer depletion, clear experimental signals  can be targeted, aiming at efficient nuclear energy release from isomers.

\begin{table}[!h]
  \centering
  \footnotesize
  \begin{tabular}{lcccc}
  \hline\hline
 $nl_j$ & $E_0$ [keV] & $S_p$ [b eV] & $S_v$ [b eV] & $S_v$ [b eV] \\
 & & & $m=3$ & $m=5$\\ 
  & & & &  \tabularnewline[-0.4cm] \hline
  & & & &  \tabularnewline[-0.4cm]
$2s_{1/2}$ & $5.20$ & $8.05 \times 10^{-4}$ & $1.14 \times 10^{-3}$ & $1.14 \times 10^{-3}$  \\
$2p_{1/2}$ & $5.19$ & $7.85 \times 10^{-5}$ & $1.35 \times 10^{-3}$ & $3.34 \times 10^{-3}$ \\
$2p_{3/2}$ & $6.02$ & $1.25 \times 10^{-5}$ & $4.21 \times 10^{-4}$ & $7.61 \times 10^{-3}$  \\

  & & & & \tabularnewline[-0.4cm] \hline\hline
  \end{tabular}
  \caption{NEEC resonance strength for isomer depletion of $^{152m}$Eu for both plane wave  $S_p$ and vortex  $S_v$ electron beams. We assume $\zeta = p_z$ and $\zeta b = 1$ and consider two values of the topological charge $m=3,\, 5$.}
  \label{table:eu152v}
\end{table}

Let us now finally turn to the magnitude of isomer depletion for the $^{93m}$Mo isomer. The isomers can be obtained in nuclear reactions such as $^{93}$Nb(p,n)$^{93m}$Mo \cite{GunstPRL2014} or  $^{7}$Li$(^{90}$Zr, p3n)$^{93}$Mo \cite{ChiaraNature2018}. Since the resonance condition for electron recombination needs to be fulfilled in the rest frame of the nucleus,  the ion preparation is equally important to the vortex electron beam generation. 
The required ion charge state breeding, storage and cooling requires for instance a storage ring or an electron beam ion trap in conjunction with a radioactive beam facility.  Isomeric beams have been successfully produced and stored at facilities such as the GSI Darmstadt  \cite{LitvinovNIMPRB2013, GrieserEPJST2012, DickelPLB2015}. At a storage ring the condition  $\zeta = p_z$ could be easily fulfilled by exploiting the Lorentz boost of the ions. A dedicated electron vortex beam setup needs to be designed in order to fulfill all experimental requirements for isomer production, resonance condition match and dynamical control of vortex beam properties. 

 Considering the most efficient capture orbital $2p_{3/2}$ and  topological charge $m=5$, the  NEEC resonance strength reaches the value $\sim 1$ b eV. 
 In order to obtain a reaction rate per ion, we multiply this value by the vortex beam flux. We assume here the generic flux of $10^{24}$ cm$^{-2}$ s$^{-1}$ eV$^{-1}$ \cite{BecheUM2017, ReimerBook2008}. Variations around this figure depend on the exact continuum electron energy required by the resonance condition. Electron energies below 1 keV will diminish the electron density, such that additional compression would be required, whereas much larger energies can even enhance the flux we are considering. The NEEC reaction rate per ion reaches the value of approx. 1 s$^{-1}$. Compared to the natural decay of the isomer (halflife 6.8 h), this represents an enhancement of approx. 4 orders of magnitude for the isomer depletion rate.

Isomer depletion is a very desirable goal in view of the current search for energy storage solutions \cite{Koningstein2014,PrelasBook2016}.
However, the potential of dynamically controlled vortex beams extends farther than that. We anticipate new opportunities in nuclear physics, where projectile beams, starting for instance from protons, neutrons or muons with reshaped wave fronts \cite{Luski2021,zhao2021decay} would enhance and dynamically control nuclear reactions. The beam angular momentum is ideal to specifically select reaction channels according to the final-state spin. This would enable for instance the targeted production of isotopes or isomers for medical applications \cite{Habs2011,Pan2020} or the search for dark matter \cite{Pospelov2020}. Thus,  nuclear physics and engineering will benefit from the new opportunities raised by vortex beams with intense flux and dynamical control of beam parameters.  
In addition, the experimental methods described above, combining controlled atomic beams (be they electrons or other particles) with tailored external handles, will offer a unique perspective for the interplay between the nucleus and its surrounding electronic shells, with potential also for chemistry and molecular physics applications.

\section{Methods}
In order to derive the NEEC rate for vortex electron beams, we relate to the plane wave results in  Refs.~\cite{PalffyPRA2006, PalffyPRA2007, GunstPOP2015} and expand the continuum electronic wave function into partial waves of definite angular momentum. To specify the lateral position of the ion with regard to the central axis of the incident electron beam, the impact parameter ${\bf{b}}$ is introduced \cite{BliokhPR2017, SerboPRA2015}. The NEEC rate can be written as
\begin{equation}
  Y_{neec}^{i \rightarrow g} = \int \mathcal{Y}_{neec}^{i \rightarrow g}({\bf{p}}, {\bf{k}}) a_{\zeta m}({\bf{p}}_{\bot}) a^{*}_{\zeta m}({\bf{k}}_{\bot}) e^{i ({\bf{k}}_{\bot}-{\bf{p}}_{\bot}) {\bf{b}}} \frac{d^2 {\bf{p}}_{\bot}}{(2 \pi)^2} \frac{d^2 {\bf{k}}_{\bot}}{(2 \pi)^2} d^2 {\bf{b}},
\end{equation}
where $\mathcal{Y}_{neec}^{i \rightarrow g}({\bf{p}}, {\bf{k}})$ is the squared transition amplitude for incoming momenta ${\bf{p}}$ and ${\bf{k}}$. We restrict the impact parameter region to $|{\bf{b}}| \leqslant b$.  The NEEC rate takes then the from
\begin{equation}
  Y_{neec}^{i \rightarrow g} = \frac{b^2}{4 \pi} \int_0^{2\pi} \!\!\! \int_0^{2\pi} \frac{d \alpha_p}{2 \pi} \frac{d \alpha_k}{2 \pi} e^{i m(\alpha_p - \alpha_k)} \mathcal{Y}_{neec}^{i \rightarrow g}({\bf{p}}, {\bf{k}}) {}_{0}F_1(2;u)/\Gamma(2) ,
\end{equation}
with the condition $|{\bf{p}}_{\bot}| = |{\bf{k}}_{\bot}| = \zeta$, and the two polar angles $\alpha_p$ and $\alpha_k$ spanning the interval $[0,2\pi)$. The notation ${}_{0}F_1$ stands for the confluent hypergeometric limit function, $u = - b^2 \zeta^2 \left[ 1-\cos{(\alpha_k - \alpha_p)} \right]/2$, and $\Gamma(2)$ is the Gamma function. 

The remaining factor $\mathcal{Y}_{neec}^{i \rightarrow g}({\bf{p}}, {\bf{k}})$ can be related to the plane-wave NEEC amplitude calculated in Refs.~\cite{PalffyPRA2006, PalffyPRA2007}
\begin{eqnarray}
  \mathcal{Y}_{neec}^{i \rightarrow g}({\bf{p}}, {\bf{k}}) &=& \frac{2 \pi (4 \pi) (2 J_g + 1) \rho_i}{2 (2 I_i + 1) (2J_i + 1) (2 j_g + 1)} \\ 
  & & \times \sum_{M_i s} \sum_{M_g m_g}  \langle I_g M_g, n_g \kappa_g m_g | H_N | I _i M_i, {\bf{p}} s\rangle \langle I_g M_g, n_g \kappa_g m_g | H_N | I _i M_i, {\bf{k}} s \rangle^{\dagger}, \nonumber
\end{eqnarray}
where $H_N$ is the electron-nucleus interaction Hamiltonian, $J_i$ is the total angular momentum of the initial electronic configuration of the ion, $J_g$  the total angular momentum of the final electronic configuration of the ion after NEEC, and $\rho_i$  the initial density of continuum electron states, respectively. The nuclear initial state (final state after NEEC) is determined by the total angular momentum $I_i$ ($I_g$) and its projection $M_i$ ($M_g$). The bound electron in the capture orbital is determined by the principal quantum number $n_g$,  the Dirac angular momentum quantum number $\kappa_g$, and projection $m_g$ of the angular momentum. Furthermore, $j_g$ is the total angular momentum of the bound electron in the capture orbital. The calculation of the electron matrix elements requires the continuum electron states with definite asymptotic momentum ${\bf{p}}$ (or ${\bf{k}}$) and spin projection $s$ to be expanded in terms of partial waves $| \varepsilon \kappa m_j \rangle$ \cite{PalffyPRA2006, PalffyPRA2007}, where $\varepsilon$ is the kinetic energy, $\kappa$ is the Dirac angular momentum quantum number, and $m_j$ is the projection of the total angular momentum $j$. The contribution of each partial wave is given by \cite{PalffyPRA2006, PalffyPRA2007}
\begin{eqnarray}
  & & \langle I_g M_g, n_g \kappa_g m_g | H_N | I _i M_i, \varepsilon \kappa m_j \rangle \nonumber\\
  &=& \frac{1}{R_0^{L+2}}  \sum_{M} (-1)^{I_g + M_i + L + M + m_j + 3j_g} \left[ \frac{4\pi (2j_g + 1)}{(2L+1)^3} \right]^{1/2} \langle I_g||\mathcal{Q}_L|| I_i\rangle \nonumber\\
  & & \times ~C(I_i~I_g~L; -M_i~M_g~M) ~C(j~J_g~L; -m_j~m_g~-M) ~C(j_g~L~j; \frac{1}{2}~0~\frac{1}{2}) R^{(E)}_{L, \kappa_g, \kappa}, \label{el}
\end{eqnarray}
for transitions of electric multipolarity $L$, and
\begin{eqnarray}
   & & \langle I_g M_g, n_g \kappa_g m_g | H_N | I _i M_i, \varepsilon \kappa m_j \rangle \nonumber\\
   &=& \sum_{M} (-1)^{I_i - M_i + M + j - L - 1/2} \left[ \frac{4\pi (2j + 1)}{L^2 (2L+1)^2} \right]^{1/2} \langle I_g||\mathcal{M}_L|| I_i\rangle (\kappa + \kappa_g) \nonumber\\
  & & \times  ~C(j~L~j_g; m~-M~m_g) ~C(I_g~I_i~L; M_d~-M_i~M) \left(\begin{array}{ccc} j_g & j & L \\ \frac{1}{2} & -\frac{1}{2} & 0\end{array}\right) R^{(M)}_{L, \kappa_g, \kappa},
\end{eqnarray}
for  transitions of magnetic multipolarity $L$. Here $\langle I_g||\mathcal{Q}_L|| I_i\rangle$ and $\langle I_g||\mathcal{M}_L|| I_i\rangle$ 
are the reduced matrix elements of the electric and magnetic multipole moments, respectively. The are connected to the reduced nuclear transition probabilities by the expression $\mathcal{B}\uparrow(E/M L)=\langle I_g||\mathcal{Q}_L/\mathcal{M}_L|| I_i\rangle/(2I_i+1)$. Furthermore, $R_0$ in Eq.~\eqref{el} denotes the nuclear radius. The radial integrals $R^{(E)}_{L, \kappa_g, \kappa}$ and $R^{(M)}_{L, \kappa_g, \kappa}$ for electric and magnetic multipolarities, respectively, are given in Refs.~\cite{PalffyPRA2006, PalffyPRA2007}.

With the expansion of  the continuum electronic wave function into partial waves of definite angular momentum, and the above matrix elements for each partial wave, we obtain the factor
\begin{equation}
  \mathcal{Y}_{neec}^{i \rightarrow g}({\bf{p}}, {\bf{k}}) = 4 \pi Y_a \! \sum_{\kappa, m_l} \!\! \frac{Y_b}{2l + 1} Y^{*}_{l m_l}(\theta_k, \varphi_k) Y_{l m_l} (\theta_p, \varphi_p),
\end{equation}
where $Y_{lm_l}$ stand for the spherical harmonics with quantum numbers $l$ and $m_l$. Furthermore, $\theta_{p}$  ($\theta_{k}$) and $\theta_{p}$ ($\theta_{k}$) are the polar  and azimuthal angles of the electron momentum $\bf{p}$ ($\bf{k}$) in the spherical coordinate of the ion.
For NEEC transitions of electric multipolarity $L$,
\begin{equation}
  Y_a = \frac{4 \pi^2 (2J_g + 1)}{(2J_i + 1) (2L+1)^2} \frac{1}{R_0^{2(L+2)}} \mathcal{B}\uparrow (EL) \rho_i,
\end{equation}
and
\begin{equation}
  Y_b = \left[ C(j_g~L~j; \frac{1}{2}~0~\frac{1}{2}) \right]^2 \left| R^{(E)}_{L, \kappa_g, \kappa} \right|^2.
  \label{aprate}
\end{equation}
For NEEC transitions of magnetic multipolarity $L$,
\begin{equation}
  Y_a = \frac{4 \pi^2 (2J_g + 1)}{(2J_i + 1) L^2 (2L+1)^2} \mathcal{B}\uparrow (ML) \rho_i,
\end{equation}
and
\begin{equation}
  Y_b = (2j +1) (\kappa_g + \kappa)^2 \left(\begin{array}{ccc} j_g & j & L \\ \frac{1}{2} & -\frac{1}{2} & 0\end{array}\right)^2 \left| R^{(M)}_{L, \kappa_g, \kappa} \right|^2.
  \label{rate}
\end{equation}
In the equations above,  $j$ is the total angular momentum of the continuum electron which connects with $\kappa$ via $j = |\kappa| - 1/2$.  The radial integrals
$R^{(E/M)}_{L,j_g,j}$  that enter Eqs.~(\ref{aprate}) and (\ref{rate}) are calculated numerically. We use
relativistic Coulomb-Dirac wave functions for the continuum electron and wave
functions calculated with the GRASP92 package \cite{ParpiaCPC1996} considering a
homogeneously charged nucleus for the bound electron. The finite size of the nucleus is not affecting significantly the radial wave functions. We find 
the values of $R^{(E/M)}_{L,j_g,j}$
are nearly constant whether or not we take
into account the finite size of the nucleus or we use Coulomb-Dirac radial wave functions. However, the finite size of the
nucleus has a sensitive effect on the energy levels of the bound
electron. The bound electron energy levels are calculated with GRASP92 and
include quantum electrodynamics corrections.

\section{Acknowledgements}
The authors thank I. Madan and G. M. Vanacore for fruitful discussions. SG, FC and AP acknowledge support from Google Inc. AP gratefully acknowledges the Heisenberg Program of the Deutsche  Forschungsgemeinschaft (DFG). 

\section{Author contributions}
YW and AP developed the theoretical formalism. YW performed the analytical and numerical calculations. 
SG and FC provided the input on experimental vortex beam parameters. AP conducted the project. 
All authors discussed the results and wrote the manuscript.




\end{document}